\documentclass{aa}  

\usepackage{graphicx}
\usepackage{txfonts}

\usepackage{natbib}
\usepackage{dcolumn,url,longtable}
\usepackage{amssymb}
\usepackage{color}
\usepackage{tikz}
\usetikzlibrary{shapes,arrows}

\newcolumntype{d}{D{.}{.}{-1}}

\newcommand{\um}{\mu\mathrm{m}}

\begin{document} 

  \title{Asteroid models reconstructed from the Lowell Photometric Database and WISE data}

  \author{J. \v{D}urech         \inst{1}        \and
          J. Hanu\v{s}          \inst{1}        \and
          V. Al\'i-Lagoa        \inst{2}        
          }

  \institute{Astronomical Institute, Faculty of Mathematics and Physics, Charles University, V Hole\v{s}ovi\v{c}k\'ach 2, 180\,00 Prague 8, Czech Republic\\
             \email{durech@sirrah.troja.mff.cuni.cz}
        \and
             Max-Planck-Institut f\"{u}r extraterrestrische Physik, Giessenbachstra{\ss}e, Postfach 1312, 85741 Garching, Germany
             }

  \date{Received ?; accepted ?}

  \abstract
  % context heading (optional)
  {Information about the spin state of asteroids is important for our understanding of the dynamical processes affecting them. However, spin properties of asteroids are known for only a small fraction of the whole population.}
  % aims heading (mandatory)
  {To enlarge the sample of asteroids with a known rotation state and basic shape properties, we combined sparse-in-time photometry from the Lowell Observatory Database with flux measurements from NASA's WISE satellite.}
  % methods heading (mandatory)
  {We applied the light curve inversion method to the combined data. The thermal infrared data from WISE were treated as reflected light because the shapes of thermal and visual light curves are similar enough for our purposes. While sparse data cover a wide range of geometries over many years, WISE data typically cover an interval of tens of hours, which is comparable to the typical rotation period of asteroids. The search for best-fitting models was done in the framework of the Asteroids@home distributed computing project.}
  % results heading (mandatory)
  {By processing the data for almost 75,000 asteroids, we derived unique shape models for about 900 of them. Some of them were already available in the DAMIT database and served us as a consistency check of our approach. In total, we derived new models for 662 asteroids, which significantly increased the total number of asteroids for which their rotation state and shape are known. For another 789 asteroids, we were able to determine their sidereal rotation period and estimate the ecliptic latitude of the spin axis direction. We studied the distribution of spins in the asteroid population. Apart from updating the statistics for the dependence of the distribution on asteroid size, we revealed a significant discrepancy between the number of prograde and retrograde rotators for asteroids smaller than about 10\,km.}
  % conclusions heading (optional)
  {Combining optical photometry with thermal infrared light curves is an efficient approach to obtaining new physical models of asteroids. The amount of asteroid photometry is continuously  growing and joint inversion of data from different surveys could lead to thousands of new models in the near future.}

  \keywords{Minor planets, asteroids: general, Methods: data analysis, Techniques: photometric}

  \maketitle

  \section{Introduction}

    The spin state and shape are among the basic physical characteristics of asteroids. The knowledge of these characteristics can help us to understand dynamical processes, such as collisions \citep{Bot.ea:15b}, thermal effects \citep{Vokrouhlicky2015}, and rotational disruption \citep{Wal.Jac:15}, for example,  that have been affecting the distribution of spins and
shapes in the main asteroid belt. The spin and shape properties can be reconstructed from photometric disk-integrated measurements if the target is observed at a sufficiently wide range of geometries \citep{Kaa.ea:02c}.

    The number of asteroid models reconstructed from photometry has been rapidly increasing due to the availability of a robust and fast inversion technique \citep{Kaasalainen2001a} and a growing archive of photometric data \citep{Warner2009, Oszkiewicz2011}. The reliability of models derived from photometry was confirmed by independent methods \citep{Mar.ea:06, Kel.ea:10, Durech2011}. The main motivation for reconstructing more asteroid models is (apart from detailed studies of individual targets of particular interest) the possibility to reveal how the spin states and shapes are distributed in the asteroid population and which physical processes affect them \citep[see][for example]{Sli.ea:09, Hanus2011, Hanus2013c, Hanus2018a, Kim.ea:14}. We aim to improve the statistics of the distribution of spins and shapes in the asteroid population. New models can be derived not only by collecting more new observations, but also just by processing archival photometric observation of large surveys. This data-mining approach was used by \citet{Hanus2011, Hanus2013a, Hanus2016a} and \citet{Durech2016}, for example. 

    In terms of quantity, the largest sources of photometric data are sparse-in-time measurements obtained by large sky surveys. While the inversion of sparse data is essentially the same as the inversion of dense light curves, a unique solution of the inverse problem can be found only for a small fraction of asteroids due to the high noise in the data. In anticipation of the publication of more accurate data from Gaia or LSST, we have already processed the available data, namely photometry from astrometric surveys compiled in the Lowell Observatory photometric database \citep{Oszkiewicz2011, Durech2016}. As the next step, in this paper we derive hundreds of new asteroid models using the Lowell photometric database in combination with thermal infrared data observed by the Wide-field Infrared Survey Explorer \citep[WISE,][]{Wright2010} and retrieved, vetted and archived in the framework of the NEOWISE survey \citep{Mainzer2011a}.

  \section{Method}
  \label{sec:method}

    When processing the data, we proceeded the same way as \citet{Hanus2011}. Then we applied the light curve inversion method of \citet{Kaasalainen2001a} to the data sets described below (Sec.~\ref{sec:convex}). The crucial task was to select only reliable solutions of the inverse problem.

  \subsection{Input data}

    We combined two photometric data sources: (i) sparse-in-time brightness measurements in V filter from the Lowell Observatory photometric database and (ii) thermal infrared data from the NEOWISE survey.

    The Lowell Observatory photometric database consists of sparse-in-time photometry from 11 large sky surveys that was re-calibrated to remove the most prominent systematic trends \citep{Oszkiewicz2011, Bowell2014a}. The data are available for more than 300,000 asteroids, with the number of points per object ranging from tens to hundreds. The accuracy of photometry is around 0.15--0.2\,mag. Most of the measurements are from the years 2000--2012.

    The second source of data was the WISE catalog \citep{Wright2010, Mainzer2011a}. The observations were made in four bands at 3.4, 4.6, 12, and 22\,$\um$, usually referred to as W1, W2, W3, and W4 data.  We retrieved the Level 1b data from the WISE All-sky database by querying the IRSA/IPAC service for each NEOWISE detection reported to and vetted by the Minor Planet Center. We rejected all measurements potentially affected by artefact contamination as flagged by the WISE Moving Object Pipeline Subsystem \citep[WMOPS,][]{Cutri2012}. Only measurements with quality flags A, B, or C, and artefact flags 0, p, or P were accepted. More details about these criteria can be found in \cite{AliLagoa2017} and references therein. 
 
    Thermal infrared data of asteroids such as W3 and W4 are typically used to derive their thermophysical properties by means of a thermophysical model \citep[see the review by][for example]{Delbo2015}. Although it would be, in principle, possible to search for a unique model using the photometry and thermal data in a fully thermophysical approach -- with the method of \citet{Durech2017c}, for example -- this would be, in practice, extremely time consuming when dealing with a large number of objects. Instead, we used another approach that we tested in \citet{Durech2015c}, where we treated the WISE thermal fluxes as reflected light. More specifically, we took the data as relative light curves assuming that the shape of a visual light curve is not very different from a light curve at thermal wavelengths under the same observing geometry. This is true for main-belt asteroids with typical values of thermal inertia (tens to hundreds SI units) and rotation period (several hours or longer). 
    
\begin{figure*}
     \begin{center}
      \resizebox{\hsize}{!}{\includegraphics{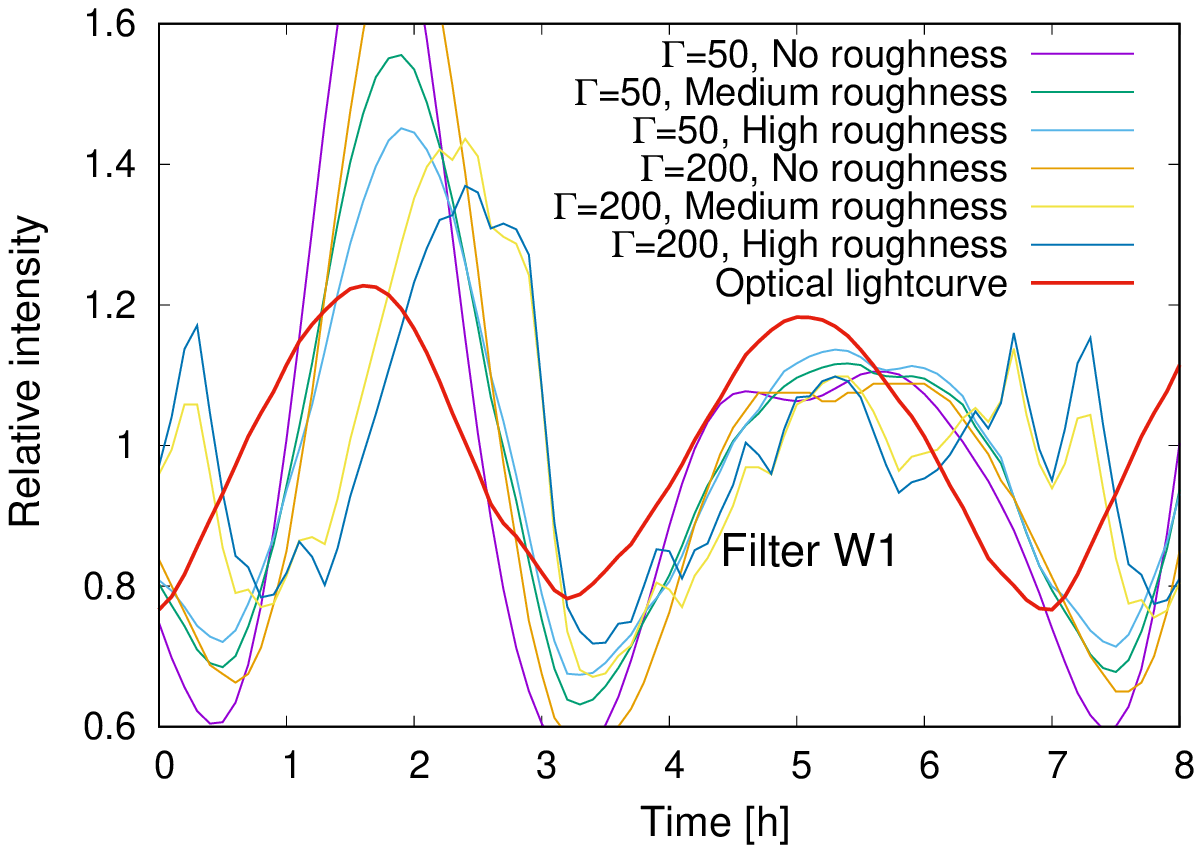}\includegraphics{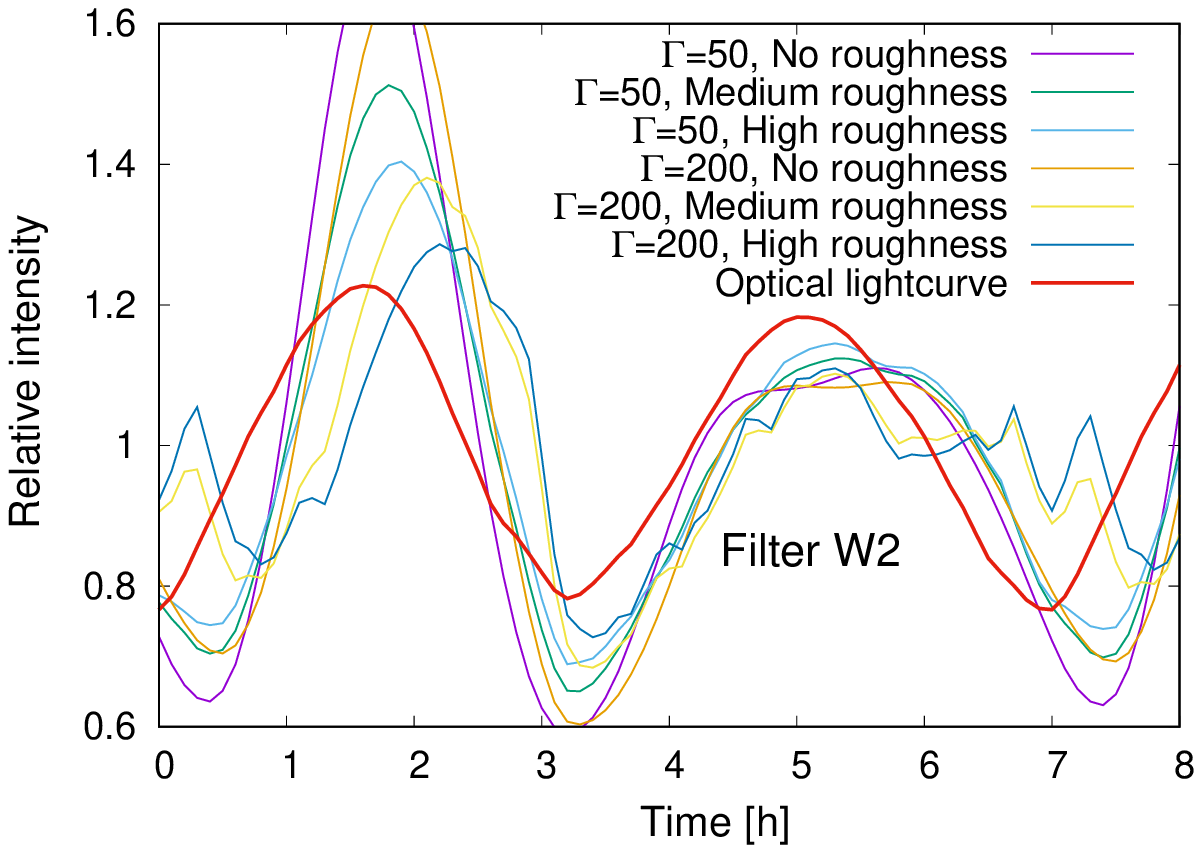}}\\
      \resizebox{\hsize}{!}{\includegraphics{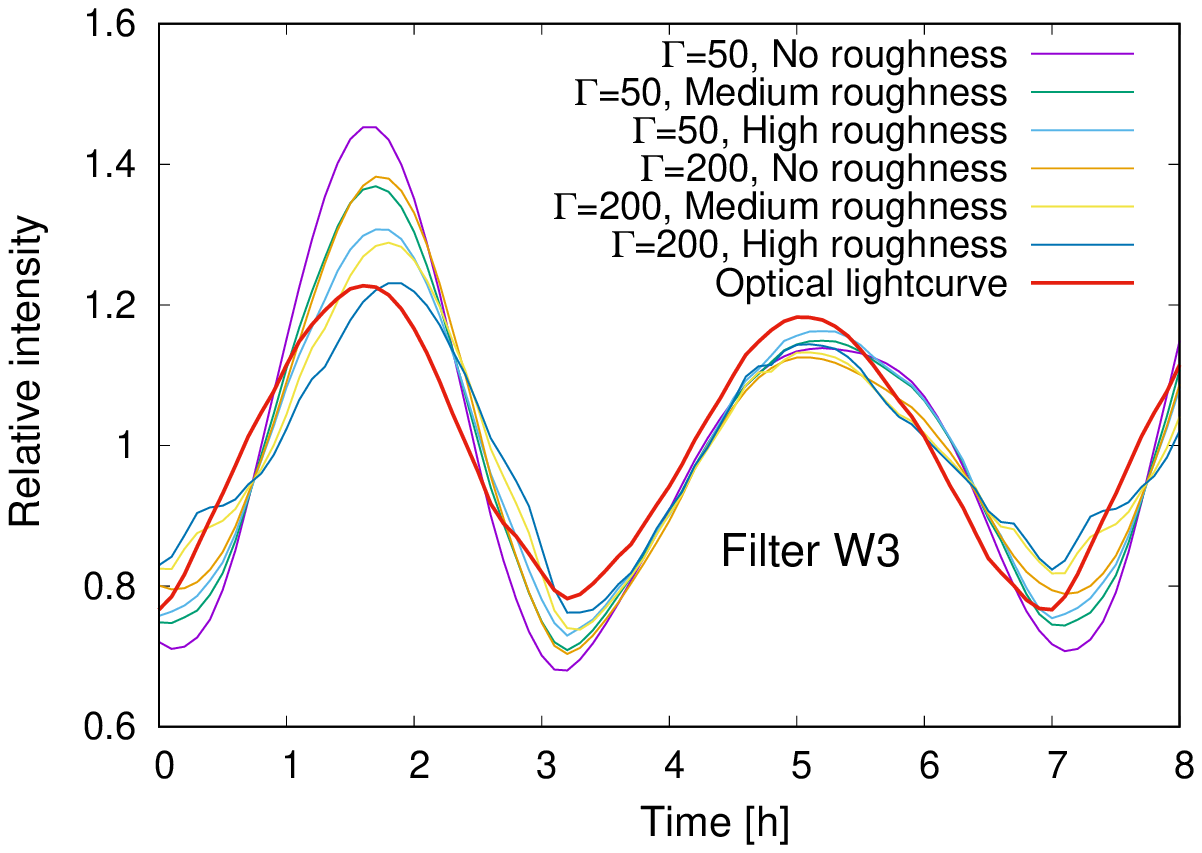}\includegraphics{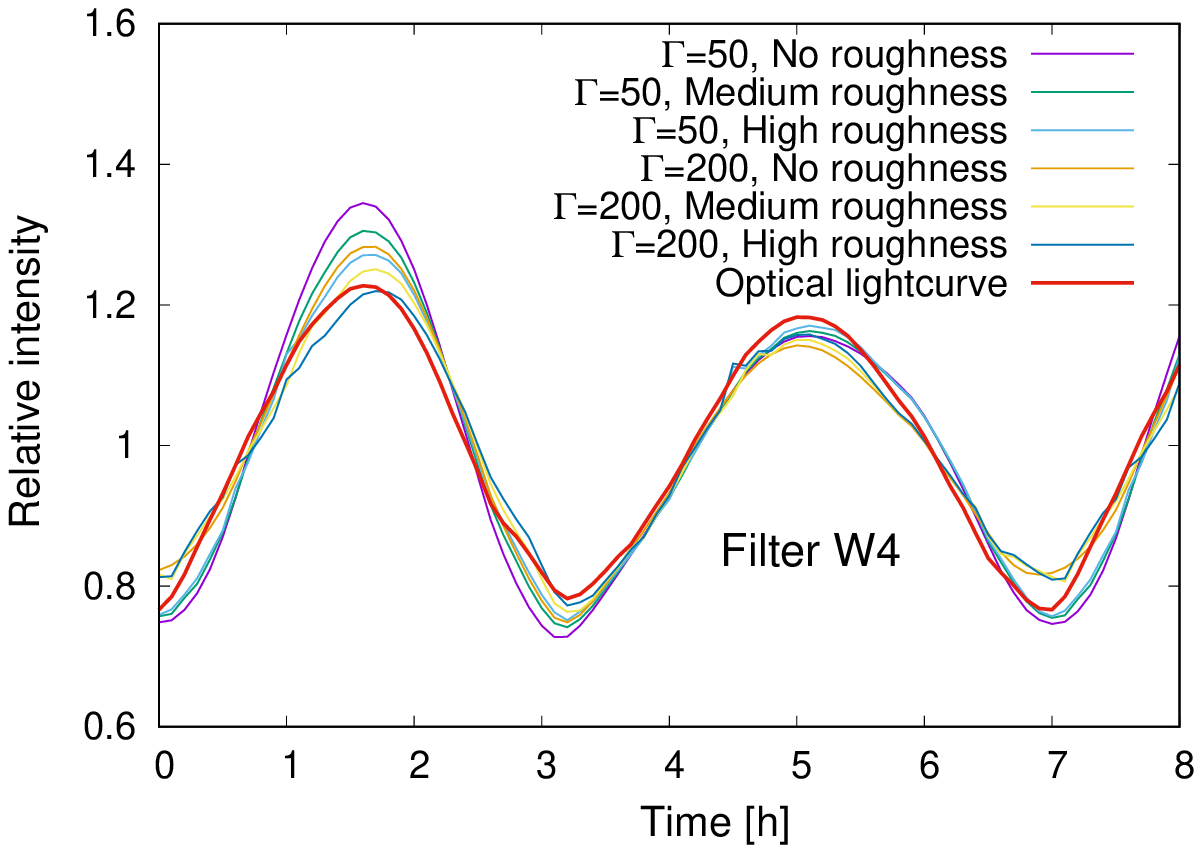}}\\
      \caption{\label{fig:reflected_vs_thermal} Thermal (in WISE channels W1, W2, W3 and W4) and optical light curves generated for a shape model of asteroid 15 Eunomia and different values of thermal inertia $\Gamma$ and macroscopic surface roughness $\overline\theta$.}
     \end{center}
    \end{figure*}

    To further support the validity of the assumption of the similarity between the optical and thermal light curves, we generated thermal light curves for several configurations; these are compared in Fig.~\ref{fig:reflected_vs_thermal} to the optical light curve generated by a standard ray-tracing algorithm. Our observing configuration and thermal properties correspond to typical values expected for a main-belt asteroid. Without loss of generality, we selected a shape model of asteroid (15)~Eunomia derived by \citet{Nathues2005} as our referenced shape model. The observing geometry was the following: the asteroid was located at a heliocentric distance of 2.5\,AU with the phase angle of 20$^{\circ}$ (this corresponds to a typical WISE observation of a main-belt asteroid), the sidereal rotation period was set to seven hours and we observed the asteroid equator-on. To generate the thermal light curve, we used the implementation of \cite{Delbo2004} and \citet{Delbo2007a} of the thermophysical model (TPM) developed by \cite{Spencer1989}, \cite{Spencer1990}, \citet{Lagerros1996, Lagerros1997, Lagerros1998}, and \cite{Emery1998}. A detailed description of the model can be found in \citet{Hanus2015a,Hanus2018b}. We used two values of thermal inertia as input for the TPM: 50 and 200\,J\,m$^{-2}$\,s$^{-1/2}$\,K$^{-1}$. Such values are typical for main-belt asteroids \citep{Hanus2018b}. Moreover, for each thermal inertia value, we ran the TPM with three different degrees of the macroscopic roughness model $\overline\theta$. We parametrize $\overline\theta$ by hemispherical craters with an opening angle $\gamma_\mathrm{c}$ and an areal coverage $\rho_\mathrm{c}$. Our model includes no roughness ($\gamma_\mathrm{c} = 0$, $\rho_\mathrm{c} = 0$), medium roughness (50, 0.5), and high roughness (90, 0.9). The TPM includes additional parameters that we fixed to realistic values (absolute magnitude, slope parameter, geometric visible albedo, Bond albedo). As we study only the normalized thermal light curve, the absolute size of the shape model is irrelevant. For generating the optical light curve, we used the combination of a single Lomell-Seeliger and multiple Lambertian scattering laws in the ray-tracing algorithm. 
   
    The majority of asteroid thermal infrared data from WISE was obtained in the W3 and W4 channels, where asteroid thermal emission dominates over most inertial sources. The thermal light curves in W3 and W4 filters for all combinations of $\Gamma$ and $\overline\theta$ are qualitatively consistent with the optical light curve (Fig.~\ref{fig:reflected_vs_thermal}), which justifies our use of the thermal light curves in filters W3 and W4 as if they were reflected light. We note that there are differences between the optical and thermal light curves; mostly the amplitudes of the thermal light curves are slightly larger than of the optical light curve. On the other hand, the positions of minima and maxima are consistent. As a result, the shape modeling with the thermal data treated as a reflected light should provide reliable rotation states, whereas the elongation of the shape models could be slightly overestimated. However, this effect seems to be negligible because when we compared shape elongations of our new models with those of models derived from only visual photometry, the difference was small and in the opposite direction (see Sect.~\ref{sec:models_comparison}). 
    
\begin{figure*}
      \begin{center}
        \includegraphics[width=0.32\textwidth]{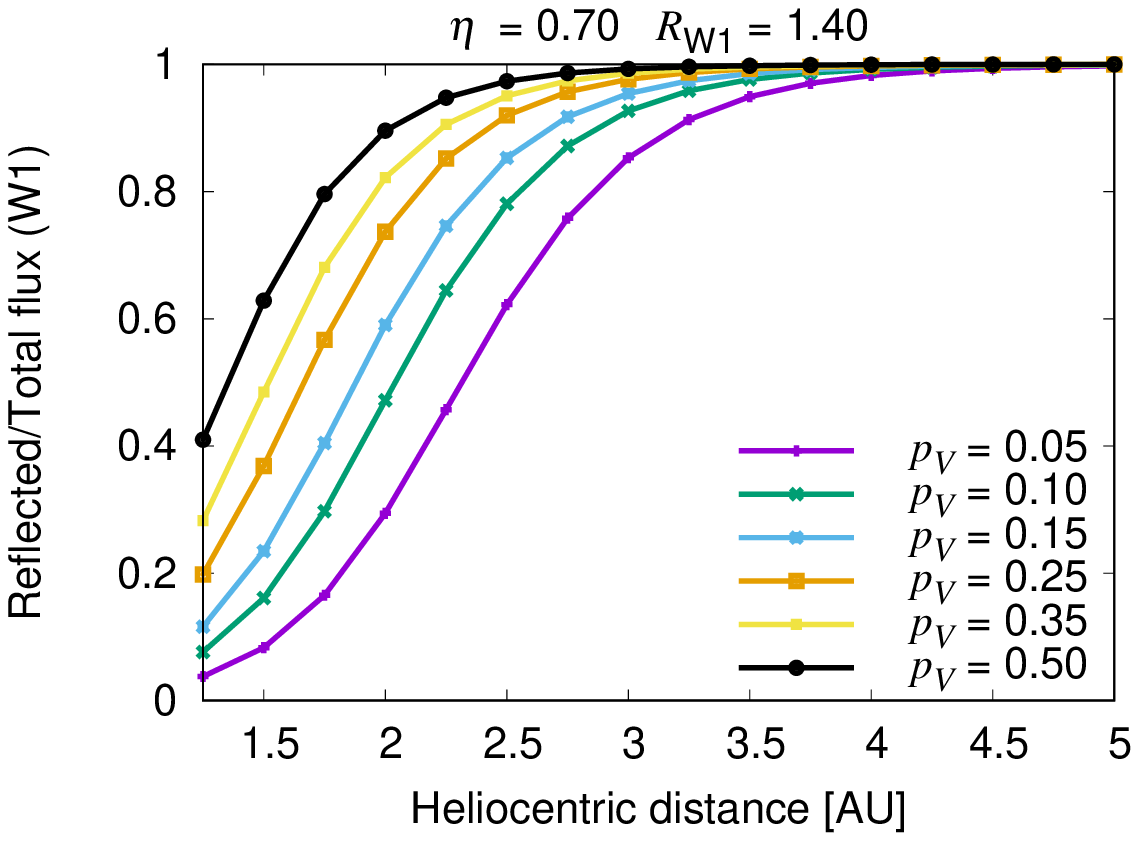} 
        \includegraphics[width=0.32\textwidth]{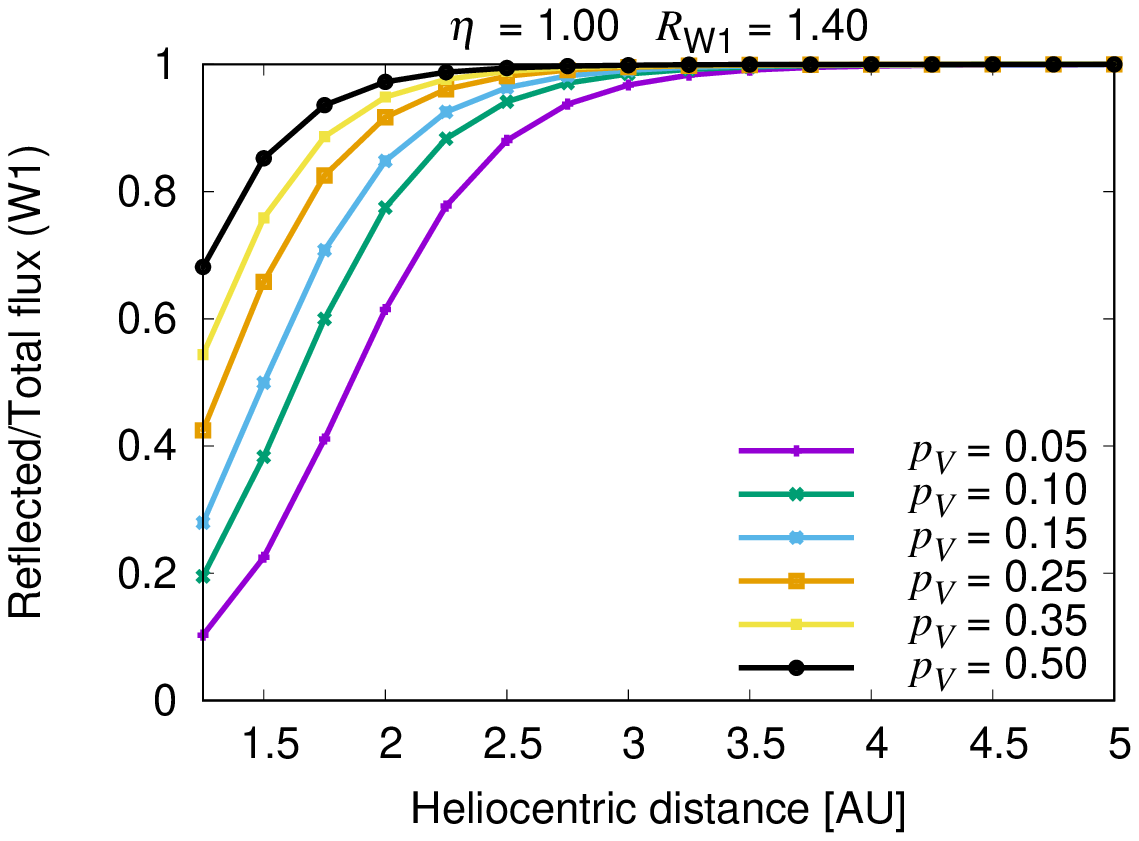}
        \includegraphics[width=0.32\textwidth]{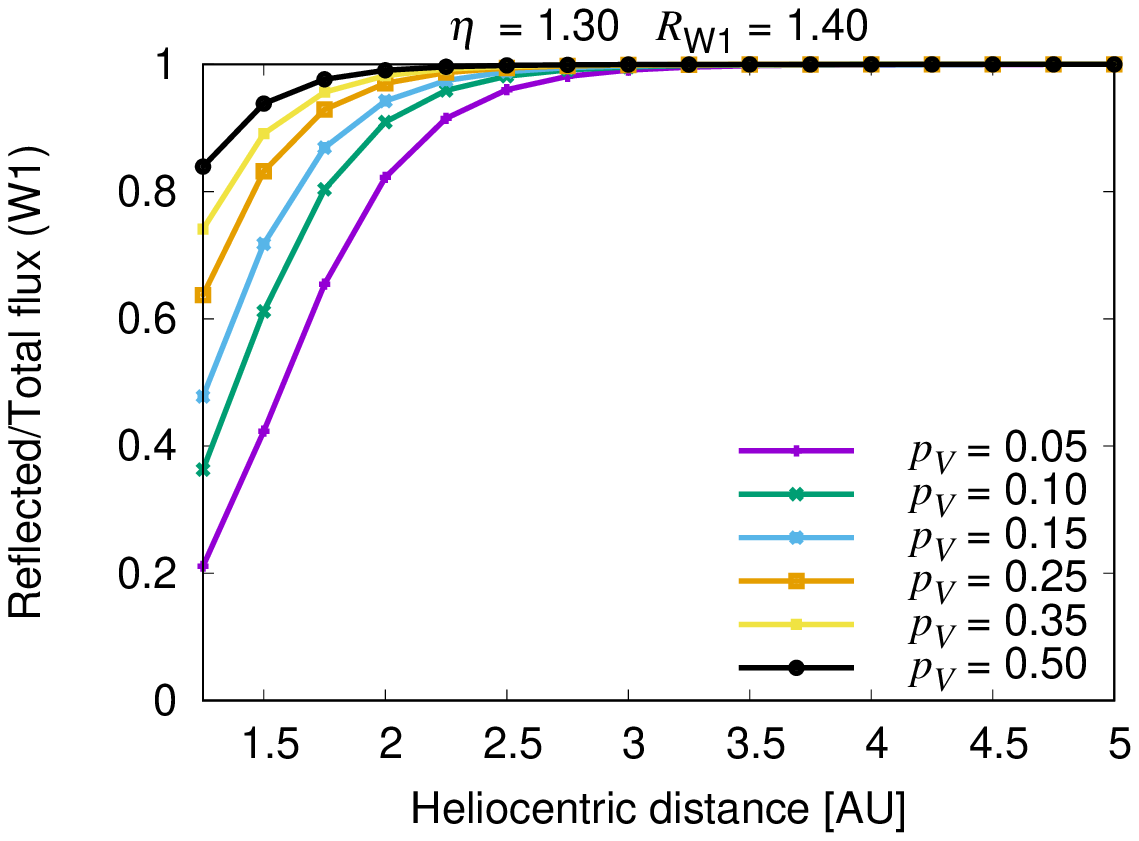}\\
        \includegraphics[width=0.32\textwidth]{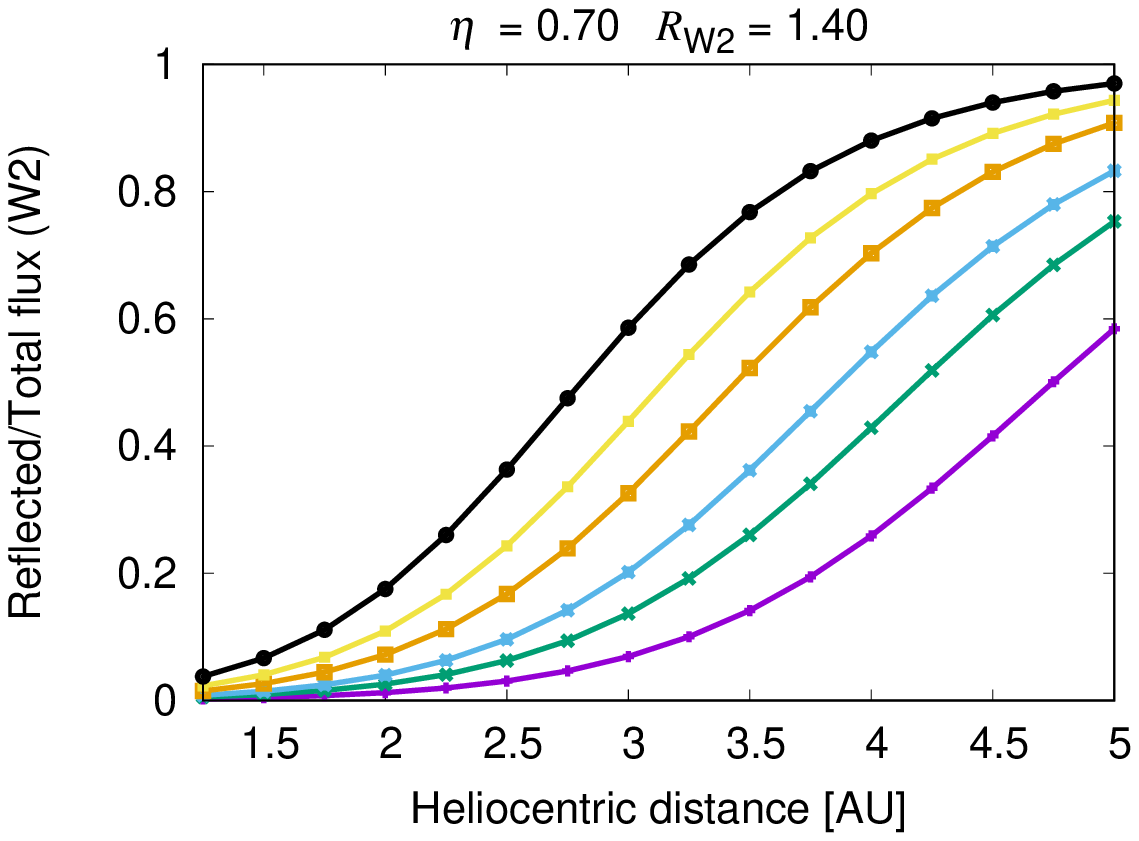}
        \includegraphics[width=0.32\textwidth]{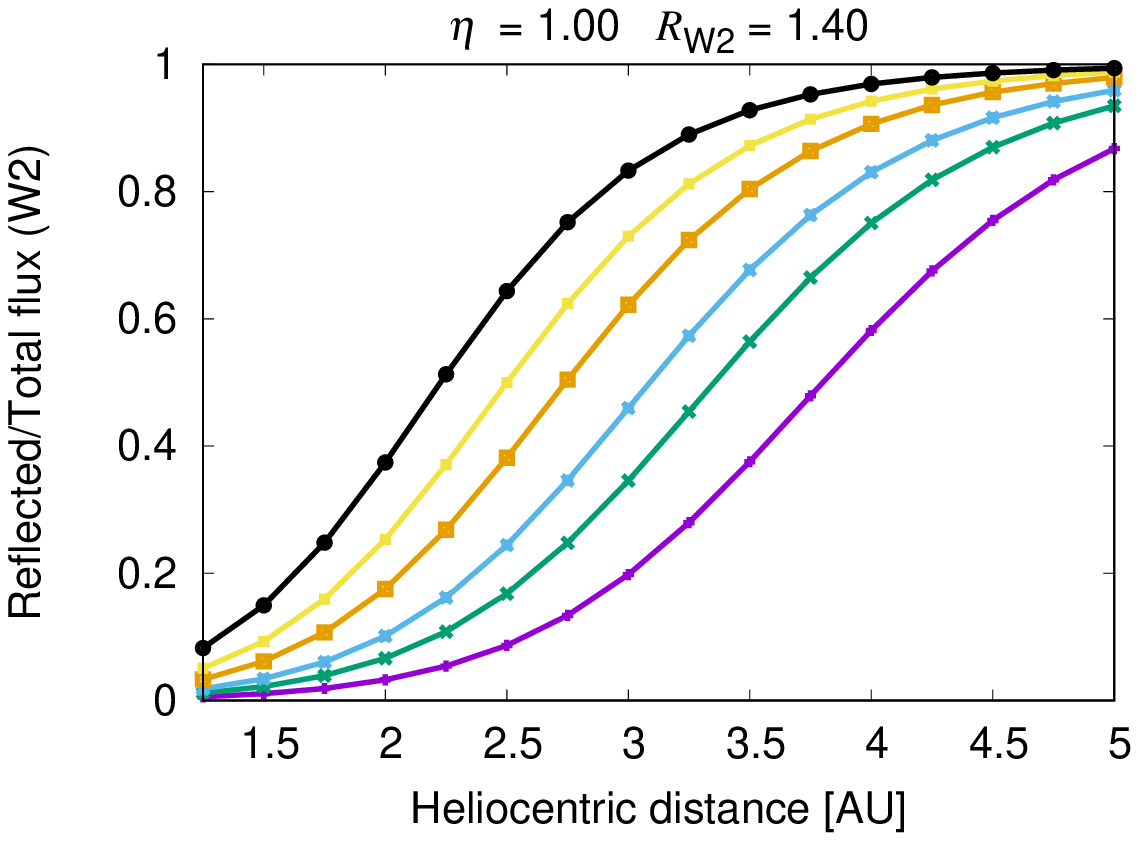}
        \includegraphics[width=0.32\textwidth]{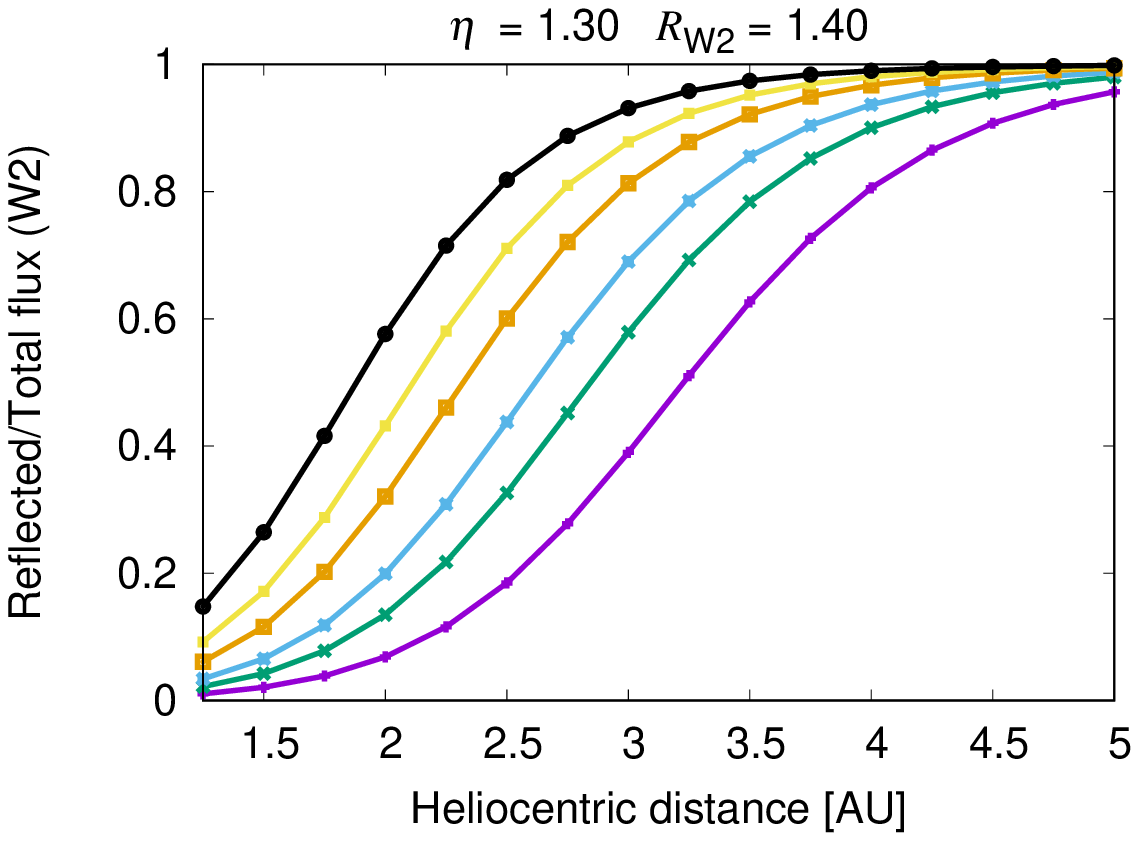}    
        \caption{Ratio of reflected to total flux in the WISE short-wavelength bands W1 (top row) and W2 (bottom row) vs. heliocentric distance predicted by the NEATM for different values of visible geometric albedo ($p_V$), and beaming parameter ($\eta$) increasing from left to right. A typical value of 1.4 was chosen for the infrared relative reflectance or albedo ratio ($R_{\mathrm{X}}\equiv p_{\mathrm{X}}/p_V$, with X$=$W1,~W2; see, e.g., \cite{AliLagoa2013} for more details). All other things being equal, a higher $\eta$ means a lower surface temperature and hence a higher reflected light contribution. For main-belt asteroids, the average $\eta$-value is 1.0\label{fig:contamination} }
      \end{center}
    \end{figure*}

    The thermal light curves of main belt asteroids in filters W1 and W2 differ more from the optical light curve than those in filters W3 and W4: the relative amplitudes are often significantly larger and the minima and maxima are shifted with respect to those of the optical light curve (see Fig.~\ref{fig:reflected_vs_thermal}). Some of the thermal light curves are not smooth; this is due to the internal numerical limitations in the surface roughness implementation in the TPM code. Fortunately, real data in filter W1 are almost always dominated by the reflected component (see Fig.~\ref{fig:contamination}), meaning that the small thermal contribution is not important for the overall flux. The only exception are dark objects in the inner main belt, but higher-albedo igneous asteroids are more numerous in this region. The situation in filter W2 is more complicated as is illustrated in Fig.~\ref{fig:contamination}. The relative contributions of the thermal and reflected components to the total observed flux depend on the surface temperature distribution, which is a complicated function of the heliocentric distance, geometric visible albedo, shape, rotation state, and so on. Depending on these parameters, the thermal component in W2 can range from a few to almost one hundred percent of the total flux. For the most common cases this fraction is between 30 and 70\%. Still, the thermal light curves are not too different from the optical one, so the total light curve composed from reflected and emitted parts should not differ substantially from the optical light curve. We note that most of the asteroids for which observations are available in the W2 filter also have data in W3 and W4 filters, which diminishes the role of the W2 data in the shape modeling. Moreover, the amount of data pertaining to observations in W1 and/or W2 filters represents only a few percent of the whole WISE All-sky catalog. 
    
    We also tested other values of thermal inertia, different input shape models, and different observing geometries. In all cases, we obtained a qualitative agreement with the conclusions above based on the shape model of Eugenia.

  \subsection{Convex models}\label{sec:convex}

    To find a physical model that fits the photometric data, we represented the shape by a convex polyhedron and used the light curve inversion method of \citet{Kaasalainen2001a}. We assumed that there was no albedo variation over the surface -- this assumption is necessary for the mathematical uniqueness of the shape solution and is generally accepted because asteroids visited by spacecraft show only small surface albedo variations. The rotation state was described by the sidereal rotation period $P$ and the ecliptic coordinates $(\lambda, \beta)$ of the spin axis direction (i.e., the pole). The search in the $P, \lambda, \beta$ parameters space was done the same way as in \cite{Durech2016}: we scanned the 2--100\,h interval of periods. For each trial period, we ran the shape optimization with ten initial pole directions. These time-consuming computations were performed using the distributed computing project Asteroids@home \citep{Durech2015b}. The whole interval of periods was divided into smaller intervals with roughly similar computing requirements and these tasks were distributed among volunteers participating in the project. Once they returned all the results, we combined them into the final periodogram. Subsequently, we identified the globally best-fitting solution and verified its reliability.
  
  \subsection{Ellipsoids}

    Similarly to \citet{Durech2016}, we also used an additional shape parametrization to find the correct rotation period, namely a model of a triaxial geometrically scattering ellipsoid. Given the poor photometric accuracy of the data, this simple model fits the data well enough to be efficiently used for the period search. The shape was described by only two parameters -- semiaxes ratios $a/c$ and $b/c$. Because the brightness can be computed analytically in this case \citep{Kaasalainen2007, Ostro1984}, this approach is approximately one hundred times faster than modeling the shape as a convex polyhedron. Moreover, by setting $a > b > c$, where $c$ is the axis of rotation, the model automatically fulfills the condition that the rotation is around the shortest axis with the maximum moment of inertia. This condition cannot be easily fulfilled during the period search with convex shapes, only checked ex post, so in many cases convex models that formally fit the data should be rejected because they are unrealistically elongated along the rotation axis and such rotation would not be stable. However, the three-dimensional (3D) shape reconstruction and inertia check is done only for the best-fitting model. Therefore, in practice, ellipsoidal models are more efficient in finding the correct rotation period because convex models can formally fit the data with an incorrect period with a nonphysical shape due to their flexibility. After finding the best rotation period with the ellipsoidal model, we switched back to convex shape representation for the subsequent pole search.  

  \subsection{Tests}
  \label{sec:tests}
  
    Having the periodograms for each asteroid, the critical task was to decide if the formally best solution with the lowest $\chi^2$ is indeed the correct solution, that is, whether the minimum in the $\chi^2$ is significant, or just a random fluctuation. To decide this, we performed a number of tests in almost exactly the same way as in \citet{Durech2016} when processing only Lowell data. The only difference was that instead of having a fixed threshold of 5\% for the $\chi^2$ increase $\chi^2_\mathrm{tr} = 1.05\,\chi^2_\mathrm{min}$, we computed the acceptance level for each asteroid individually according to the formula $\chi^2_\mathrm{tr} = (1 + 0.5 \sqrt{2/\nu})\,\chi^2_\mathrm{min}$. This is nothing more than an empirical prescription to take into account the number of measurements ($\nu$ is the number of degrees of freedom, i.e., the difference between the number of data points and the number of parameters). In our case, we have three parameters for the spin state, three parameters to fit the photometric phase function, and $(n+1)^2$ shape parameters with convex shapes or two parameters for ellipsoids. With convex shapes, $n$ is the degree and order of the spherical harmonics expansion \citep{Kaasalainen2001b}. The empirical formula for $\chi^2_\mathrm{tr}$ is related to the fact that the $\chi^2$ distribution with $\nu$ degrees of freedom has mean $\nu$ and variance $2\nu$, so the formal $1\sigma$ interval for a normalized $\chi^2/\nu$ is $1 \pm \sqrt{2 / \nu}$. The multiplicative factor $1/2$ is an adjustment without which the threshold would be too high and the number of unique models too low. For comparison, the 5\% level used in our previous analysis now corresponds to $\nu = 200$.

    Here we summarize the steps that we took to select the final models. These steps are essentially the same as those of our previous analysis in \citet{Durech2016} so we leave out the details; they are shown in a flow chart in Fig.~\ref{fig:flow_chart}.

    \begin{enumerate}
      \item The period interval 2--100\,h was scanned independently with convex models with two shape resolutions $n = 3$ and $n = 6$ and with ellipsoids.
      \item For each periodogram, we found the period with the lowest $\chi^2_\mathrm{min}$. We defined this period as unique if all other periods outside the uncertainty interval had $\chi^2$ higher than the threshold $\chi^2_\mathrm{tr}$ defined above.
      \item The unique periods for two different resolutions of the convex model had to be the same within the error interval.
      \item If the unique period was longer than 50\,h, we checked if there is no deeper minimum for periods that were longer than the original interval of 2--100\,h: we ran the period search again with an interval of 100--1000\,h. 
      \item If there were more than two pole solutions defined again by the $\chi^2_\mathrm{tr}$, we reported such models as partial if they had constrained $\beta$ (see Sect.~\ref{sec:partial_models}).
      \item If there were two possible poles, the difference in $\beta$ had to be smaller than $50^\circ$ and the difference in $\lambda$ 120--240$^\circ$ -- this corresponds to the $\lambda \pm 180^\circ$ ambiguity for observations restricted to regions near the ecliptic plane \citep{Kaasalainen2006}. 
      \item Check if the rotation is along the shortest axis.
      \item Visual check of the fit, residuals, and the shape.
    \end{enumerate}

    %\tikzstyle{decision} = [diamond, draw, fill=blue!20, text width=4.5em, text badly centered, node distance=2.5cm, inner sep=0pt]
    \tikzstyle{block} = [rectangle, draw, fill=blue!20, text width=5em, text centered, rounded corners, minimum height=4em]
    \tikzstyle{line} = [draw, thick, color=black, -latex']
    %\tikzstyle{cloud} = [draw, ellipse,fill=red!20, node distance=2.5cm, minimum height=2em]

    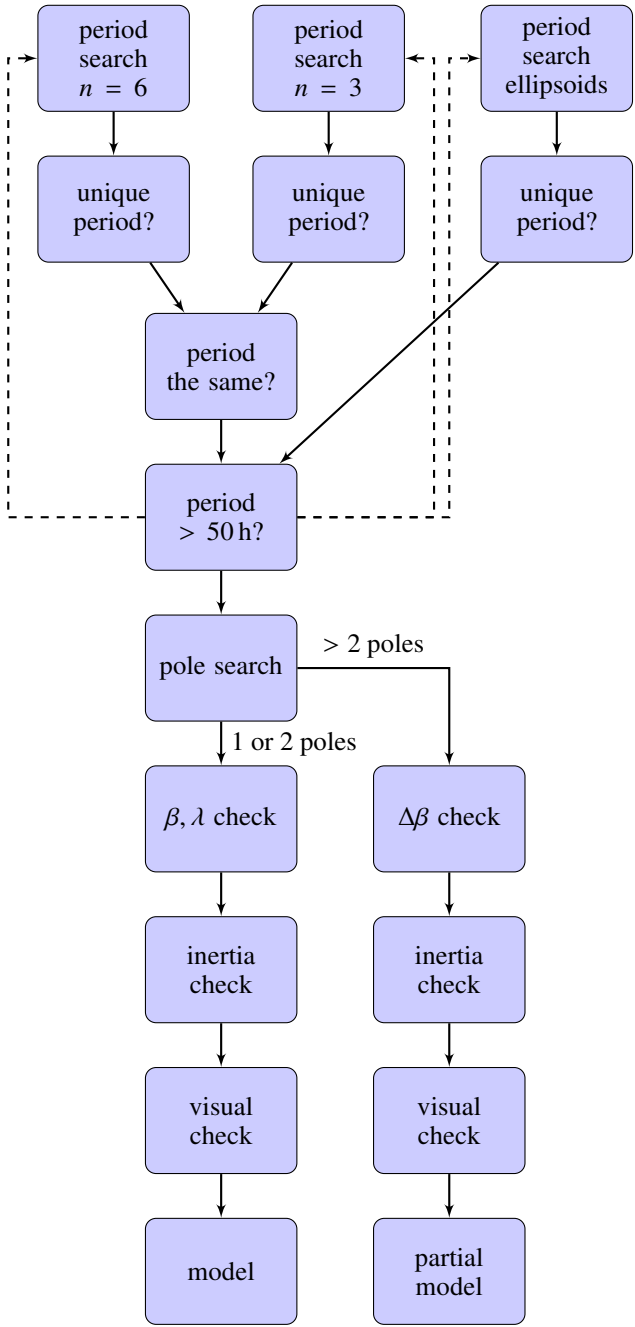
\begin{figure}
     \begin{center}
      
      \begin{tikzpicture}[scale=2, node distance = 2cm, auto]
       \node (none1) [text width=0cm] {};
       \node [block, below left of=none1] (period_n6) {period search $n = 6$};
       \node [block, below right of=none1] (period_n3) {period search $n = 3$};
       \node [right of=period_n3, text width=0cm] (none3) {};
       \node [block, right of=period_n3, node distance = 3cm] (period_ell) {period search ellipsoids};
       \node [block, below of=period_n6] (unique_n6) {unique period?};
       \node [block, below of=period_n3] (unique_n3) {unique period?};
       \node [block, below of=period_ell] (unique_ell) {unique period?};
       \node (none2) [text width=1cm, below of=none1] {};
       \node [block, below of=none2, node distance=3.5cm] (period_the_same) {period the same?};
       \node [block, below of=period_the_same] (period_gt50) {period $> 50\,$h?};
       \node [block, below of=period_gt50] (pole) {pole search};
       \node [block, below of=pole] (beta_lambda) {$\beta, \lambda$ check};
       \node [block, below of=beta_lambda] (inertia) {inertia check};
       \node [block, below of=inertia] (visual) {visual check};
       \node [block, below of=visual] (model) {model}; 
       \node [block, right of=model, node distance = 3cm] (partial_model) {partial model};
       \node [block, right of=beta_lambda, node distance = 3cm] (Delta_beta) {$\Delta \beta$ check};
       \node [block, below of=Delta_beta] (inertia_partial) {inertia check};
       \node [block, below of=inertia_partial] (visual_partial) {visual check};

       \path [line] (period_n6) -- (unique_n6);
       \path [line] (period_n3) -- (unique_n3);
       \path [line] (period_ell) -- (unique_ell);
       \path [line] (unique_n3) -- (period_the_same);
       \path [line] (unique_n6) -- (period_the_same);
       \path [line] (period_the_same) -- (period_gt50);
       \path [line] (period_gt50) -- (pole);
       \path [line] (unique_ell) -- (period_gt50);
       \path [line, dashed] (period_gt50) -- ++(1.5,0) |- (period_ell);
       \path [line] (pole) -- node {1 or 2 poles} (beta_lambda);
       \path [line] (beta_lambda) -- (inertia);
       \path [line] (inertia) -- (visual);
       \path [line] (visual) -- (model);
       \path [line] (pole) -| node [near start, color=black] {$>2$ poles} (Delta_beta);
       \path [line] (Delta_beta) -- (inertia_partial);
       \path [line] (inertia_partial) -- (visual_partial);              
       \path [line] (visual_partial) -- (partial_model);                
       %\path [line] (decide) -- node [, color=black] {no}(stop);
       \path [line,dashed] (period_gt50) --  ++ (1.4,0) |- (period_n3);
       \path [line,dashed] (period_gt50) --  ++ (-1.4,0) |- (period_n6);
      \end{tikzpicture}
      \caption{Steps in the processing pipeline that selects reliable solutions only.}
      \label{fig:flow_chart}
     \end{center}
    \end{figure}
  
  \section{Results}
  \label{sec:results}

    By processing data for all $\sim 75,000$ asteroids for which we had enough observations, we ended up with 908 reconstructed unique models. Out of these, 246 were already known from inversion of other data and served as an independent check of reliability and error estimation. The efficiency is low because of poor photometric accuracy of sparse photometry, but still significantly higher than when processing Lowell sparse data alone \citep{Dur.ea:16}.
  
    \subsection{Comparison with independent models}
    \label{sec:models_comparison}

      The models of 246 asteroids that were already reconstructed from other data -- not necessarily fully independent because many of them were based on Lowell sparse data \citep{Durech2016} -- and made available through the Database of Asteroid Models from Inversion Techniques \citep[DAMIT,][]{Durech2010} were used for various tests. We used this subset mainly to check the frequency of false positive results. Out of 246 models, five had periods that were slightly different from published values. The two different periods corresponded to different local minima; the relative difference between periods was of the order of only $10^{-4}$, but in most cases two slightly different periods led to largely different pole directions. Periods for four other asteroids were completely different. In the remaining 237 cases, the period was determined correctly (or at least in agreement with the DAMIT value) and for such cases the difference between the poles was mainly less than $30^\circ$, with only a few cases having a pole difference up to $60^\circ$. The distribution of pole differences was similar to that presented by \citet{Durech2016}. The mean pole difference was $12^\circ$ and the median value was $9^\circ$. We also compared the semiaxis ratios $a/b$ and $b/c$ (computed from a dynamically equivalent ellipsoid) of our models and those in DAMIT. The mean value of $(a/b)_\text{DAMIT} / (a/b)_\text{our}$ was about 1.06 with a standard deviation of 0.13. For $(b/c)_\text{DAMIT} / (b/c)_\text{our}$, the mean value was also 1.06, while the standard deviation was higher: 0.18. Therefore, on average, our models are slightly less elongated than their counterparts in DAMIT. 

      This test showed us that, with the current setup, the majority of models we derive are ``correct'' in the sense that they agree with models based on different data sets often containing mainly dense light curves. The number of false positive solutions is a few percent. We expect that the number of incorrect period/pole solutions among the new models in the following section is about the same, that is, a few percent.

      \begin{figure*}[t]
       \begin{center}
        \includegraphics[width=\textwidth]{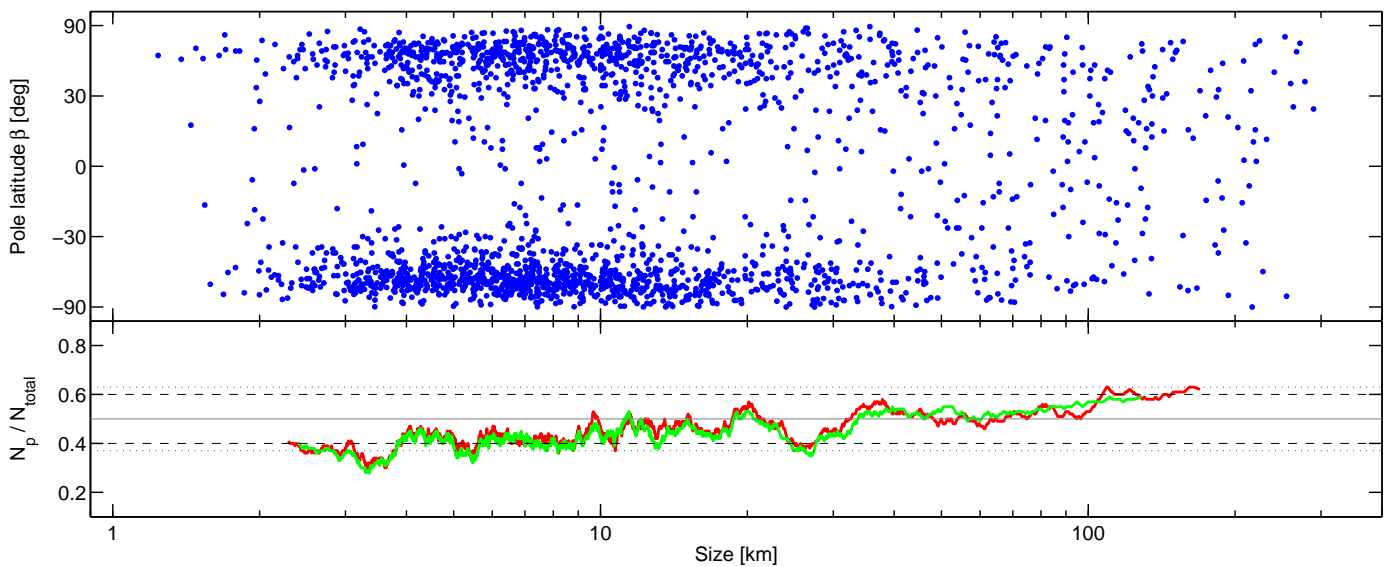}
        \caption{The distribution of pole latitudes $\beta$ for all full and partial models of main-belt asteroids. The bottom panel shows the running-box (over 100 asteroids) fraction of prograde $(\beta > 0\deg)$ models (red curve) and also the fraction of models with $\beta > 30\deg$ (green curve). The dashed and dotted lines show the 5\% and 1\% significance levels at which the null hypothesis that the prograde and retrograde asteroids are equally frequent can be rejected.}
        \label{fig:beta_vs_size}
       \end{center}
      \end{figure*}

    \subsection{New models}

      In total, we derived new models for 662 asteroids (169 using convex shape period search, 513 using ellipsoids, and 20 overlapping). These models and their parameters are listed in Table~\ref{tab:models}. All models are available in DAMIT, from where not only the shape and spin can be downloaded but also all data points that were used for the inversion. 

      We compared the derived sidereal rotation periods $P$ with those reported in the Asteroid Lightcurve Database (LCDB) of \citet{Warner2009}; version from November 12, 2017. In most cases, they agreed, which can be taken as another independent verification of the reliability of the model. In some cases however, the periods were different and we checked again if this was likely due to an erroneous model or an incorrect period in the LCDB. In only one case was it clear that our model was wrong -- we rejected asteroid (227)~Philosophia from the list of our results because our period was not consistent with dense light curves \citep{Mar.ea:18}. For other cases that were not consistent with LCDB, we checked the LCDB entries and sometimes concluded that the LCDB period is likely wrong because it was not supported by enough quality light curves (usually the uncertainty code was $< 2$). In other cases, both the LCDB entry and our model looked right and we were not able to decide if our model or the LCDB entry was wrong -- these cases are marked with an asterisk in Table~\ref{tab:models}.

    \subsection{Partial models}
    \label{sec:partial_models}

      Apart from the models described above, we also derived 789 so-called partial models \citep{Hanus2011}. These are asteroids for which the rotation period was determined uniquely but for which there were more than two possible pole solutions satisfying the $\chi^2_\mathrm{tr}$ criterion. Although we do not take such results as unique solutions of the inverse problem, they still carry important information about the rotation state. In these cases, there are more than two pole solutions that fit the data equally well, but are usually not distributed randomly. On the contrary, their $\beta$ values are often limited to one hemisphere, clearly distinguishing between prograde and retrograde rotation. This is a valuable constraint that can be used in the analysis of the spin axes distribution in the following section.

      These partial models are listed in Table~\ref{tab:models_partial}. Because the pole direction is not known, we list the mean pole latitude $\beta$ of all acceptable solutions and their dispersion $\Delta$ defined as $|\beta_\mathrm{max} - \beta_\mathrm{min}| / 2$. We list only such asteroids for which the $\beta$ values were limited within $50^\circ$, so $\Delta \leq 25^\circ$. We also individually checked those asteroids for which our period was different from that in LCDB. For some of them, we concluded that the LCDB period was not reliable; for others the LCDB period seemed reliable but so did our model, so we marked such inconsistent models with an asterisk in Table~\ref{tab:models_partial}. We rejected asteroid (6199)~Yoshiokayayoi because it was in strong disagreement with reliable LCDB data and was likely a false positive solution in our sample.

  \section{Spin distribution}

    The new models we derived significantly increased the total number of asteroids for which the spin orientation is known. Although there are also other sources of asteroid models and spin parameters \citep[the radar models, see, e.g.,][]{Benner2015}, we limited our analysis to models from DAMIT and the new models we derived here. DAMIT contains models for 943 asteroids (as of January 2018), so the total number of available models is now $\sim 1600$. There are other 789 partial models, which means that the total number of asteroids for which we have at least some information about the spin axis direction is $\sim 2400$.

    In what follows, we concentrate on the analysis of how the spin ecliptic latitude $\beta$ is distributed in the population. Other physical parameters like the shape or the rotation period are likely to be biased by the selection effects -- elongated asteroids are more likely to be successfully modeled than spheroidal asteroids because they have larger light curve amplitudes and their signal is not lost in the noise \citep{Mar.ea:18}. To draw any reliable conclusions about the distribution of the shapes or periods, we would need to carefully de-bias our sample, which is outside the scope of this paper. On the other hand, if there was any bias affecting $\beta$, it should be symmetric with respect to $\pm \beta$ and not dependent on the size, so we can readily examine and interpret the latitude distribution. 

    The $\beta$ values are related to the ecliptic plane. However, a more ``physical'' value is the pole obliquity $\gamma$, defined as the angle between the spin axis and the direction perpendicular to the orbital plane. The conversion between these two parameters is trivial for zero orbital inclination because in this case $\gamma = 90^\circ - \beta$ and the prograde/retrograde rotation exactly corresponds to the sign of~$\beta$. For nonzero inclination, the conversion depends also on the ecliptic longitude $\lambda$ of the pole and on the orbital elements $I$ (inclination) and $\Omega$ (longitude of the ascending node). Because $\lambda$ is not known for partial models, we assume the simple zero inclination conversion. For full models, there are often two possible pole solutions with, in general, different $\beta$ and also $\gamma$ values. Averaging $\gamma$ or $\beta$ values of two models would lead to smearing of the extreme values, so for the following plots we randomly selected one of them with the corresponding $\beta$. For partial models, the value of $\beta$ in the plots is taken as an arithmetic mean of the values for all acceptable poles. The orbital elements were taken from the AstOrb\footnote{\url{ftp://ftp.lowell.edu/pub/elgb/astorb.html}} database, the diameters were mainly from the NEOWISE database \citep{Mainzer2011a} with some values taken also from Akari \citep{Usui2011} and IRAS \citep{Tedesco2004} catalogs.

      \begin{figure*}[t]
        \begin{center}
          \includegraphics[width=\textwidth]{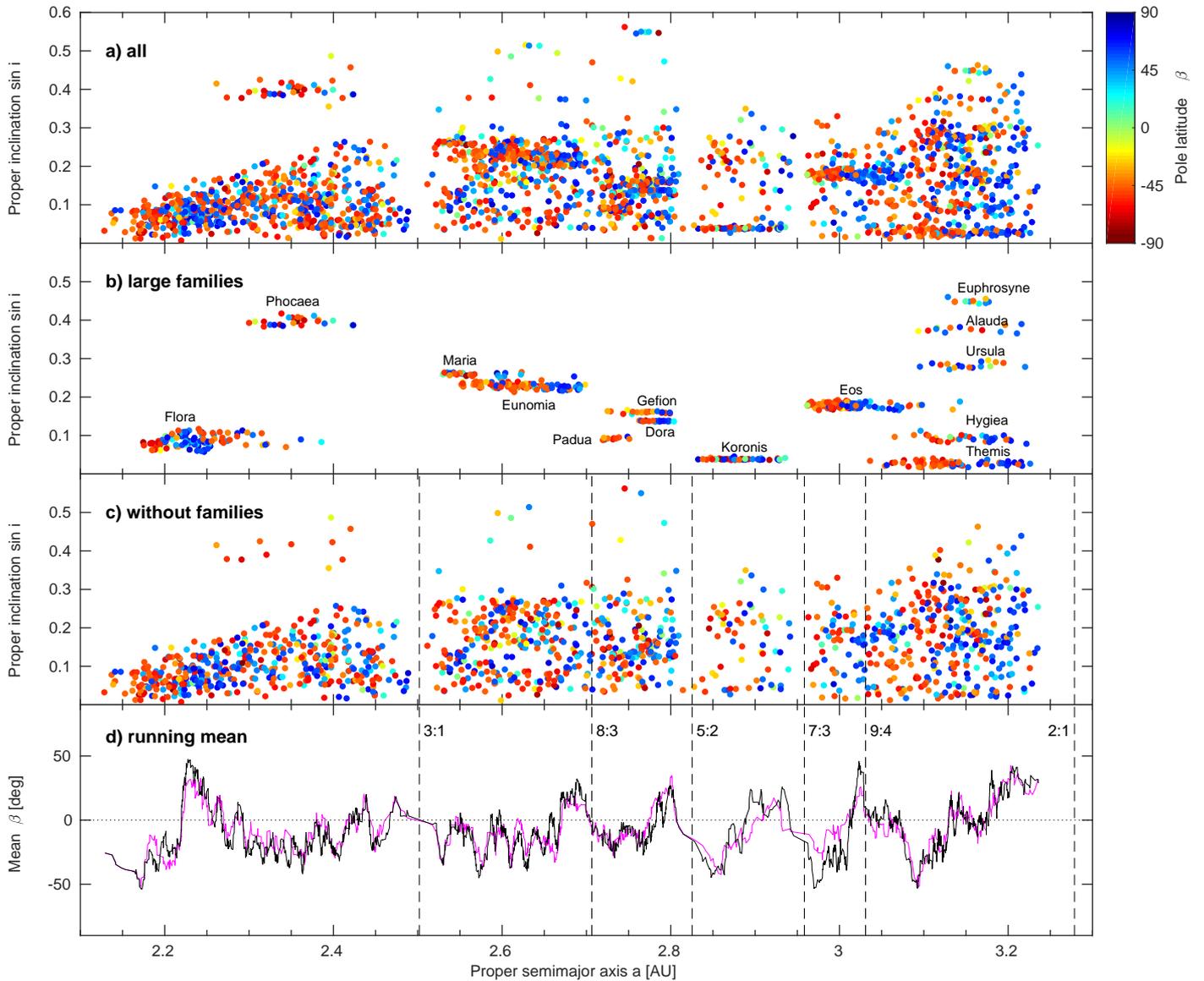}
            \caption{The distribution of pole latitude $\beta$ for all models (a), asteroids in the largest families (b), and asteroids not in families (c). The bottom panel (d) shows the running mean of $\beta$ over 20 asteroids (all asteroids -- black, without families -- magenta). The dashed vertical lines mark the strongest mean motion resonances.}
            \label{fig:beta_in_MB}
        \end{center}
      \end{figure*}

    \subsection{Pole latitude versus size}

      The distribution of the pole latitude $\beta$ for main-belt asteroids as a function of asteroid size is shown in Fig.~\ref{fig:beta_vs_size}. The distribution is strongly bimodal for asteroids smaller than 20--30\,km, which can be satisfactorily explained as an effect of a YORP-driven evolution \citep{Hanus2011, Hanus2013a}. Due to the YORP effect, small asteroids evolve towards the extreme values of obliquity. In the lower panel of Fig.~\ref{fig:beta_vs_size}, we show the fraction $N_\mathrm{p} / N_\mathrm{total}$ of the number of prograde $(\beta > 0)$ rotators in a running box over $N_\mathrm{total} = 100$ asteroids as a function of size. For asteroids larger than 100\,km, the number of prograde rotators is statistically higher ($N_\mathrm{p} = 68$, $N_\mathrm{total} = 108$, probability of the null hypothesis that $N_\mathrm{p} = N_\mathrm{total} / 2$ is $p \simeq 0.7\%$ assuming binomial distribution) in accordance with the model of \cite{Johansen2010} who suggested that the preferentially prograde rotation of large asteroids is a result of accretion of pebbles on planetesimals in a gaseous environment. On the other hand, for asteroids in the size range 1--10\,km, there is an excess of retrograde rotators ($N_\mathrm{p} = 520$, $N_\mathrm{total} = 1276$, $p \simeq 2 \times 10^{-11}$). For asteroids between 10 and 100\,km, the number of prograde and retrograde rotators is statistically the same ($N_\mathrm{p} = 441$, $N_\mathrm{total} = 919$, $p \simeq 22\%$). Because most asteroids smaller than about 30\,km  have large absolute values of $\beta$, the prograde/retrograde analysis is not sensitive to asteroids with $|\beta| < 30^\circ$. The ratio is almost the same even if we restrict ourselves to $|\beta| > 30^\circ$ where the distinction between prograde and retrograde rotation is unambiguous even for nonzero inclination. This is not true for asteroids larger than 100\,km, where for $|\beta| > 30^\circ$ we have $N_\mathrm{p} = 34$, $N_\mathrm{total} = 55$, and $p \simeq 8\%$.

      The excess of small retrograde rotators in Fig.~\ref{fig:beta_vs_size} is statistically significant, however, it is not clear if the reconstructed distribution of $\beta$ is the same as the real one. Although we are not aware of any bias in the observations or the method that could cause the asymmetry in $\pm \beta$, there could still be some nontrivial systematic effect that we have not taken into account.

    \subsection{Pole latitude distribution across the main belt}

      The distribution of pole latitude $\beta$ is not only dependent on the size, but also on the proper semimajor axis, namely on the proximity to resonances. For asteroids in a collisional family, $\beta$ depends on the relative position with respect to the center of the family. The color-coded distribution of $\beta$ across the main belt is shown in Fig.~\ref{fig:beta_in_MB}. Similarly to how the YORP effect is responsible for clustering of poles around extreme values of obliquity for small asteroids, fingerprints of the Yarkovsky effect are clearly visible in some asteroid families (e.g., Eunomia, Koronis, Eos, and Themis) where retrograde family members are concentrated to the left (smaller semimajor axis $a$) of the family center while prograde are concentrated to the right (larger $a$). This is shown as a color dichotomy in Fig.~\ref{fig:beta_in_MB}b and is in agreement with the theoretical prediction that prograde rotators migrate to a higher semimajor axis, the opposite to retrograde rotators \citep{Vokrouhlicky2015, Hanus2013c, Hanus2018a}. The excess of retrograde rotators in the right ``wing'' of the Flora family might be caused by contamination with Baptistina family members \citep{M-D.ea:05, Bot.ea:07}. However, any detailed check of family membership or a deeper study of the distribution of spins in families are outside the scope of this paper.

      Another correlation that we can see in Fig.~\ref{fig:beta_in_MB} is the one between the sense of rotation and the location with respect to the mean-motion and secular resonances. As shown by \citet{Hanus2011}, an area to the left of a resonance contains more prograde rotators because they move towards the resonance and become scattered with only a small probability of crossing the resonance. For the same reason, there are more retrograde rotators to the right of the resonance. This separation due to resonances can be seen in Fig.~\ref{fig:beta_in_MB}c. In Fig.~\ref{fig:beta_in_MB}d, we plot the running mean of $\beta$ over 20 asteroids as a function of $a$. We can see a general behavior that to the left of a resonance the mean $\beta$ is high, meaning more prograde rotators. To the right of a resonance it drops to negative values meaning retrograde rotation. At the inner end of the main belt, the $\nu_6$ resonance cuts the belt and this area contains mainly retrograde rotators. At the opposite end, the 2:1 mean-motion resonance defines the edge and this area is populated mainly by prograde rotators.

      Finally, there are also other features in Fig.~\ref{fig:beta_in_MB} that seem to be significant but for which we have no simple explanation. Namely, these are the excess of prograde rotators at $\sim 2.24\,$AU and the excess of retrograde rotators at 3.10\,AU. The former might be related to the proximity to the inner edge of the main belt and the $\nu_6$ resonance. The latter might be directly related to the 9:4 resonance, which filters out prograde rotators to the right of the resonance. All prograde rotators between 9:4 resonance and 3.1\,AU might be Eos family members, some of them not identified as belonging to the family. 

  \section{Conclusions}
  \label{sec:conclusion}

    The combination of optical photometry with thermal data turned out to be an efficient way to enlarge the sample of asteroids with shape models and spin parameters. Although the success rate of deriving a unique physical model from Lowell and WISE data is low, and the derived models are probably a very biased sample of the whole population (there is a strong bias in favor of elongated asteroids -- their light curve amplitude is larger and the signal is less likely to be lost in the noise than for spherical objects), the asymmetry and anisotropy of the pole latitude $\beta$ corresponding roughly to the difference between prograde and retrograde rotation seems to be significant.
  
    The potential of this kind of data mining is really huge, because apart from the continuously growing number of asteroid light curves, data from other surveys like ATLAS, PTF, Gaia, or LSST are or will become available.

    With the models we derive here, the next step could be the derivation of thermophysical parameters in the same way as by \citet{Hanus2018b}. We also plan to investigate in more detail the rotation states in collisional families.

  \begin{acknowledgements}
    We would not be able to process data for hundreds of thousands of asteroids without the help of the tens of thousands of volunteers who joined the Asteroids@home BOINC project and provided their computing resources. We greatly appreciate their contribution. The work of J\v{D} and JH was supported by the grants 15-04816S and 18-04514J of the Czech Science Foundation. VAL has received funding from the European Union's Horizon 2020 Research and Innovation Programme, under Grant Agreement no. 687378. This publication also makes use of data products from NEOWISE, which is a project of the Jet Propulsion Laboratory/California Institute of Technology, funded by the Planetary Science Division of the National Aeronautics and Space Administration.   
  \end{acknowledgements}

  \begin{appendix}
   
    \section{The list of new models}

  \onecolumn

  \longtab{1}{
    \tiny
\begin{longtable}{r l r r r r @{} d @{} d l r r r r r c}
      \caption{\label{tab:models} List of new asteroid models. For each asteroid, we list one or two pole directions in the ecliptic coordinates $(\lambda, \beta)$, the sidereal rotation period $P$, the rotation period from LCDB $P_\mathrm{LCDB}$ (if available) and its quality code $U$, the number of sparse photometric data points $N$, the number of data points in WISE bands W1--W4, and the method that was used to derive the rotation period: C -- convex inversion, E -- ellipsoids, CE -- both methods gave the same unique period. The accuracy of the sidereal rotation period $P$ is of the order of the last decimal place given. For steroids marked with an asterisk, there is an inconsistency between $P$ and $P_\mathrm{LCDB}$.}\\
      \hline\hline
      \multicolumn{2}{c}{Asteroid}              & \multicolumn{1}{c}{$\lambda_1$}       & \multicolumn{1}{c}{$\beta_1$}   & \multicolumn{1}{c}{$\lambda_2$}       & \multicolumn{1}{c}{$\beta_2$}   & \multicolumn{1}{c}{$P$}       & \multicolumn{1}{c}{$P_\mathrm{LCDB}$} & \multicolumn{1}{c}{U}   & \multicolumn{1}{c}{$N$}       & W1    & W2    & W3      & W4    & method        \\
      number    & name/designation              & \multicolumn{1}{c}{[deg]}             & \multicolumn{1}{c}{[deg]}       & \multicolumn{1}{c}{[deg]}             & \multicolumn{1}{c}{[deg]}       & \multicolumn{1}{c}{[h]}       & \multicolumn{1}{c}{[h]}                 &                       &                               &         &       &       &       &               \\
      \hline
      \endfirsthead
      \caption{continued.}\\
      \hline\hline
      \multicolumn{2}{c}{Asteroid}              & \multicolumn{1}{c}{$\lambda_1$}       & \multicolumn{1}{c}{$\beta_1$}   & \multicolumn{1}{c}{$\lambda_2$}       & \multicolumn{1}{c}{$\beta_2$}   & \multicolumn{1}{c}{$P$}       & \multicolumn{1}{c}{$P_\mathrm{LCDB}$} & \multicolumn{1}{c}{U}   & \multicolumn{1}{c}{$N$}       & W1    & W2    & W3      & W4    & method        \\
      number    & name/designation              & \multicolumn{1}{c}{[deg]}             & \multicolumn{1}{c}{[deg]}       & \multicolumn{1}{c}{[deg]}             & \multicolumn{1}{c}{[deg]}       & \multicolumn{1}{c}{[h]}       & \multicolumn{1}{c}{[h]}                 &                       &                               &         &       &       &       &               \\
      \hline
      \endhead
      \hline
      \endfoot
      %  227 & Philosophia               &    98 &    17 &   277 &     4 &  47.8663 &       52.98 &     3 &   465 &    6 &    6 &      &    7 &   E  \\		
           252 & Clementina                &   121 &    46 &   307 &    59 & 10.86190 &      10.864 &     3 &   408 &   10 &    8 &   10 &   11 &   E  \\
           373 & Melusina                  &    19 & $-48$ &   147 & $-61$ & 12.98632 &       12.97 &     3 &   489 &    9 &    8 &      &   11 &   C  \\
           439 & Ohio                      &    39 & $-54$ &   260 & $-30$ &  37.4675 &       37.46 &     3 &   387 &    8 &    8 &      &    9 &   E  \\
           458 & Hercynia                  &    74 &    14 &   262 &    32 &  21.8136 &      21.806 &     3 &   427 &   12 &   12 &      &   12 &   E  \\
           462 & Eriphyla                  &   119 &     7 &   301 &     5 &  8.65890 &       8.659 &     3 &   323 &    9 &    8 &   10 &    9 &   E  \\
           558 & Carmen                    &    77 & $-35$ &   254 & $-17$ & 11.39324 &      11.387 &     3 &   553 &   23 &   23 &   21 &   27 &   E  \\
           568 & Cheruskia                 &   141 & $ -2$ &       &       & 13.20869 &      13.209 &     3 &   360 &   13 &   13 &      &   13 &   E  \\
$\ast$\    647 & Adelgunde                 &    39 & $-68$ &   252 & $-67$ &   98.097 &      32.202 &     3 &   380 &   10 &    7 &   15 &   13 &   C  \\
\end{longtable}
  }

  \newpage

  \onecolumn
  
  \longtab{2}{
    \tiny
    \begin{longtable}{r l r r @{} d @{} d l r r r r r c}
      \caption{\label{tab:models_partial} List of new partial models. For each asteroid, we list the mean ecliptic latitude $\beta$ of the spin axis, its dispersion $\Delta$, and the meaning of other columns is the same as in Table~\ref{tab:models}.}\\
      \hline\hline
      \multicolumn{2}{c}{Asteroid}              & \multicolumn{1}{c}{$\beta$}   & \multicolumn{1}{c}{$\Delta$}    & \multicolumn{1}{c}{$P$}       & \multicolumn{1}{c}{$P_\mathrm{LCDB}$} & \multicolumn{1}{c}{U}   & \multicolumn{1}{c}{$N$}       & W1    & W2    & W3      & W4    & method        \\
      number    & name/designation              & \multicolumn{1}{c}{[deg]}     & \multicolumn{1}{c}{[deg]}       & \multicolumn{1}{c}{[h]}       & \multicolumn{1}{c}{[h]}                 &                       &                               &         &       &       &       &               \\
      \hline
      \endfirsthead
      \caption{continued.}\\
      \hline\hline
      \multicolumn{2}{c}{Asteroid}              & \multicolumn{1}{c}{$\beta$}   & \multicolumn{1}{c}{$\Delta$}    & \multicolumn{1}{c}{$P$}       & \multicolumn{1}{c}{$P_\mathrm{LCDB}$} & \multicolumn{1}{c}{U}   & \multicolumn{1}{c}{$N$}       & W1    & W2    & W3      & W4    & method        \\
      number    & name/designation              & \multicolumn{1}{c}{[deg]}     & \multicolumn{1}{c}{[deg]}       & \multicolumn{1}{c}{[h]}       & \multicolumn{1}{c}{[h]}                 &                       &                               &         &       &       &       &               \\
      \hline
      \endhead
      \hline
      \endfoot
                  133 & Cyrene                    & $-35$ &   7 & 12.70858 &      12.708 &     3 &   408 &      &    7 &    8 &    7 &   E  \\
             251 & Sophia                    & $-54$ &  10 &  20.2221 &      20.216 &     3 &   438 &   20 &   18 &   20 &   20 &   E  \\
             256 & Walpurga                  &    59 &  16 & 16.66015 &      16.664 &     3 &   480 &   10 &   10 &    8 &   11 &   E  \\
$\ast$\      304 & Olga                      &    52 &  18 &  19.2606 &       18.36 &     3 &   435 &   12 &   12 &      &   14 &   E  \\
$\ast$\      426 & Hippo                     & $-33$ &  17 &   67.504 &        34.3 &     2 &   366 &    7 &    7 &      &    7 &   E  \\
             733 & Mocia                     &    53 &  16 & 11.37618 &      11.374 &     3 &   364 &    8 &    9 &      &   10 &   C  \\
             919 & Ilsebill                  & $-53$ &  18 &  5.03348 &      5.0325 &     3 &   405 &   12 &   11 &   15 &   15 &   E  \\
            1070 & Tunica                    & $-58$ &  22 &  15.5763 &        15.8 &  2$-$ &   369 &    6 &    6 &    7 &    7 &   E  \\
\end{longtable}
  }

  \end{appendix}
\end{document}